\documentclass[final,3p,times,twocolumn]{elsarticle}

\usepackage{amssymb}
\usepackage{amsmath}

\usepackage[percent]{overpic}
\usepackage{xcolor}
\journal{Advances in Water Resources}
\usepackage{subcaption}

\begin{document}

\begin{frontmatter}



\title{\textit{HydroStartML}: A combined machine learning and physics-based approach to reduce hydrological model spin-up time}

\author[label1,label2]{Louisa Pawusch}
\ead{lp9617@princeton.edu}
\affiliation[label1]{organization={University of Stuttgart, Institute for Modeling Hydraulic and Environmental Systems / Stuttgart Center for Simulation Science},
            city={Stuttgart},
            country={Germany}}
\affiliation[label2]{organization= {Visiting Student at Princeton University, Dept. of Civil and Environmental Engineering},
            city={Princeton},
            state={NJ},
            country={USA}}

\author[label1]{Stefania Scheurer}
\author[label1]{Wolfgang Nowak}
\author[label3]{Reed Maxwell}
\affiliation[label3]{organization= {Princeton University, Dept. of Civil and Environmental Engineering / High Meadows Environmental Institute / Integrated GroundWater Modeling Center},
            city={Princeton},
            state={NJ},
            country={USA}}

\title{}




\begin{abstract}
Finding the initial depth-to-water table (DTWT) configuration of a catchment is a critical challenge when simulating the hydrological cycle with integrated models, significantly impacting simulation outcomes. Traditionally, this involves iterative spin-up computations, where the model runs under constant atmospheric settings until steady-state is achieved. These so-called model spin-ups are computationally expensive, often requiring many years of simulated time, particularly when the initial DTWT configuration is far from steady state.

To accelerate the model spin-up process we developed \textit{HydroStartML}, a machine learning emulator trained on steady-state DTWT configurations across the contiguous United States. \textit{HydroStartML} predicts, based on available data like conductivity and surface slopes, a DTWT configuration of the respective watershed, which can be used as an initial DTWT.

Our results show that initializing spin-up computations with \textit{HydroStartML} predictions leads to faster convergence than with other initial configurations like spatially constant DTWTs. The emulator accurately predicts configurations close to steady state, even for terrain configurations not seen in training, and allows especially significant reductions in computational spin-up effort in regions with deep DTWTs.
This work opens the door for hybrid approaches that blend machine learning and traditional simulation, enhancing predictive accuracy and efficiency in hydrology for improving water resource management and understanding complex environmental interactions.
\end{abstract}




\end{frontmatter}



\section{Introduction}
Knowledge about the available water resources on our planet is more important than ever.
Climate change has a significant impact on the hydrological cycle itself, which could potentially severely impact the available water resources \citep{Yang2021Hydrological}.
The freshwater resources, land-atmosphere feedbacks, and ecosystems change \citep{gronewold2021tug, dirmeyer2012evidence}, threatening the biodiversity \citep{sala2000global, mantyka2015climate} by depressing population sizes, for example, due to an increased frequency of extreme climatic events \citep{van2010changes}.

This trend is especially alarming when considering that the human water consumption is one of the more important mechanisms that intensify hydrological drought and likely is a major factor affecting their intensity and frequency in the future \citep{wada2013human}.
This is due to the almost quadrupling of the global population in the last $100$ years, combined with rising standards of living and the increase of per capita food demand \citep{vorosmarty2005fresh}, which leads to a concerning spread of groundwater depletion \citep{alley2002flow}.
A projection from the United Nations World Water Development report states that nearly $57\%$ of the global population will suffer from clean water scarcity by the year $2050$, while in $2018$, this had already concerned $47\%$ of the global population \citep{UN2018WaterDev}.

Since groundwater makes up for $99\%$ of the liquid fresh water on our planet \citep{shiklomanov1998world}, and groundwater levels are sensitively impacted by changing climate conditions \citep{gurdak2007climate, allen2004groundwater}, hydrological models and simulations can serve as powerful tools to manage this situation.
Information about the availability of liquid fresh water in the form of groundwater is valuable, even though it can be especially hard to gain due to the lack of exhaustive and finely resolved information on groundwater levels, e.g. through groundwater wells. 
Instead, one option to obtain information on groundwater availability is large-scale, physics-based hydrological simulation.
To assess e.g. local depletion, these large-scale models must also have a sufficiently high resolution, which makes them computationally even more complex.

Hydrological models have advanced to complex, integrated models simulating the terrestrial hydrological cycle, while integrating the natural processes that drive these systems and that we understand increasingly well \citep[a review of recent advances can be found in][]{Brookfield2023Advances}.
But any transient hydrological model requires initial conditions, which can consist of spatially varying fields that describe the surface water and energy states at the start of a simulation \citep{rodell2005evaluation}. 
Specifically, the behavior of the model and all subsequent predictions can depend sensitively on the initial depth-to-water table (DTWT) configuration \citep{SeckSpin2015, AjamiAssessing2014}.
The depth-to-water table describes the distance between the surface and the underground groundwater table.
In absence of precise and dense DTWT data, a hydrological simulation is usually initialized with a DTWT configuration that is at equilibrium \citep[e.g.][]{Maxwell2015HighResolution}.
Since the system of equations that describe the hydrological cycle are often times highly nonlinear, this equilibrium or steady state cannot be found by directly inverting the system.

Therefore, in order to obtain a meaningful initial condition, so-called spin-up computations (SUCs) are conducted.
This means applying a constant (meteoro-) hydrological forcing of averaged, constant atmospheric settings of a singular or several years, until the system equilibrates \citep{kollet2008capturing}.
Whether the system reached equilibrium is often judged through observations of the overall water stored in the saturated and unsaturated zone in the respective basin, as convergence of the change of storage during SUC timesteps towards zero implies that the system reached a steady state.

\begin{figure}
    \centering
    \includegraphics[width=8cm]{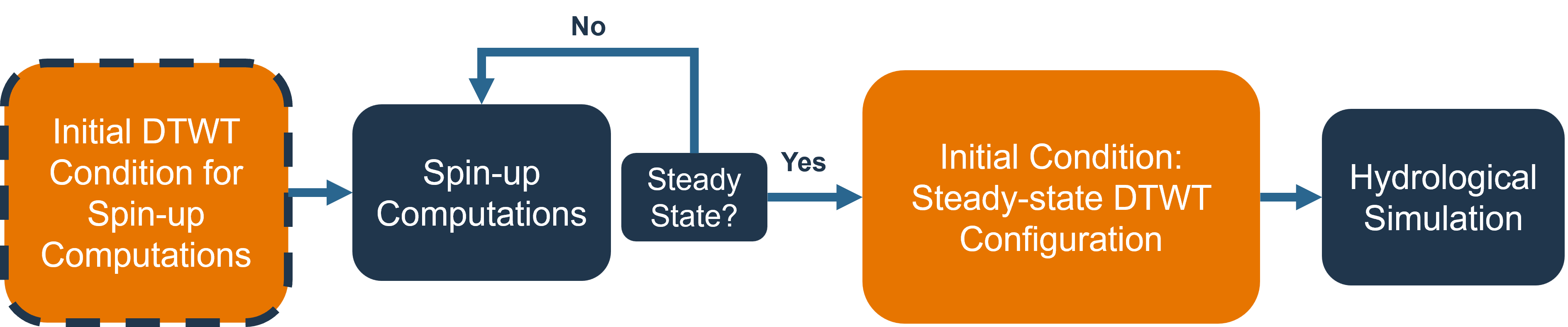}
    \caption{Typically, spin-up computations rely on an initial condition for the depth-to-water table (DTWT), and provide an initial condition at steady-state for the subsequent hydrological simulations.}
    \label{fig:flowchart_motiv}
\end{figure}

But even SUCs need an initial condition.
They are often initialized with a constant DTWT across the entire watershed \citep[e.g.][]{SeckSpin2015, Yang2023CONUS2, Erdal2019Value}, which can be understood as a single height representing the best guess for the initial DTWT for SUCs.
This initial condition converges towards the actual steady-state conditions during the subsequent SUCs.
The resulting steady-state is then used as the initial condition for any subsequent hydrological simulation.
A flowchart of this general approach for finding the initial condition for hydrological simulations is shown in Fig.~\ref{fig:flowchart_motiv}.

The remaining challenge is that these SUCs are highly resource-intensive and time-consuming, which is especially the case for large-scale models at high resolutions.
Additionally, when restarting a hydrological simulation with modified parametric values, users frequently re-run these expensive SUCs to find the new steady-state configuration.
Since small changes necessitate a huge computational effort, large-scale hydrological models are often considered highly inefficient and inflexible.

The issue of the high computational effort for SUCs has been addressed in several publications.
E.g. \citep{ajami2014reducing} discusses a hybrid approach that is based on a combination of model simulations and an empirical DTWT function.
In this application, the changes in the subsurface water storage and the DTWT during SUCs resemble an exponential decay, which was exploited by correcting the change in DTWT after a short initial spin-up period with an additional factor that stemmed from a well-parametrized exponential curve.
This approach allowed a reduction in simulation time by $50\%$.

Alternatively, some approaches reduce the computational spin-up effort by considering simplified models during SUCs. 
E.g. in \citep{Erdal2019Value}, several different simplifications of the hydrological large scale model allowed a reduction of the effort by more than two thirds.
First, a steady-state configuration, calculated with a simple two-dimensional groundwater model was applied.
In a second approach, this configuration was combined with vertical profiles in the unsaturated zone, and in a third effort, a dynamic steady state was applied by combining transient groundwater models for the subsurface and the unsaturated zone.
A lower spatial resolution of the grid was chosen for the SUCs than in the subsequent simulation step by later interpolating the steady-state configuration to the desired resolution in \citep{rodell2005evaluation}.
This approach lead to mixed results with reduced savings of computational effort if the spin-up resolution was too small and if the atmospheric settings showed large variability at small scales.

\begin{figure}
    \centering
    \includegraphics[width=0.99\linewidth]{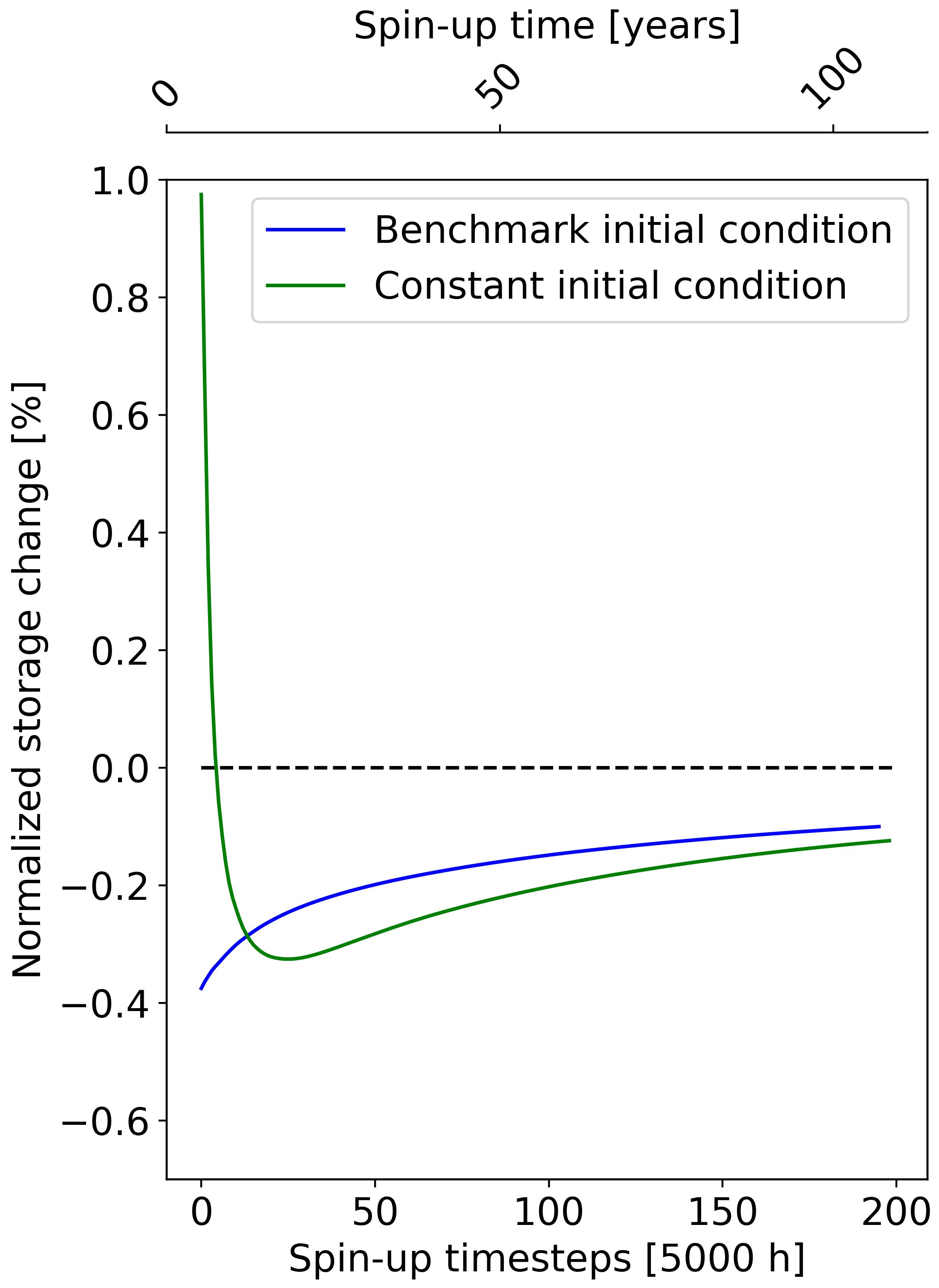}
    \caption{Convergence behavior of the normalized storage change during SUCs, initialized with different initial DTWT configurations.}
    \label{fig:Motiv_SUBehavior}
\end{figure}

The computing time of SUCs depends sensitively on their initial DTWT configurations.
Compare, for example Fig.~\ref{fig:Motiv_SUBehavior}, in which the convergence of the overall water storage in an exemplary basin during SUCs is shown for two different initial DTWT configurations.
The normalized storage change in this basin converges slowly towards zero, which can be understood as a convergence towards an equilibrium configuration.
The convergence pattern differs though, as it depends on the choice of the initial condition.

Machine learning approaches are especially useful in replicating known behaviors based upon examples.
Since the prediction quality achievable by machine learning is highly dependent on data availability, hybrid approaches that combine machine learning with physics-based models are gaining traction.
So-called emulators are methods capable of generating results similar to those of physics-based models. 
They have found application in a broad spectrum of topics in recent times \citep[e.g.][]{bennett2024spatio, agarwal2024accelerating, maxwell2021physics, tran2021development}.
The hydrological research community could benefit from such an emulator that generates steady-state DTWT configurations similar to those that would require expensive SUCs.
Initializing the required, subsequent SUCs with this steady-state DTWT as a well-informed initial DTWT configuration is a promising approach to reduce the computational effort for SUCs.

In this work, we develop the machine learning groundwater emulator \textit{HydroStartML} that decreases the computational effort of hydrological simulations on unseen terrains by faster SUCs by finding better suitable initial conditions for SUCs.
To find these suitable initial conditions for SUCs, we use existing data in the form of previously determined steady-state DTWT configurations.
Such data may be available as the result of previous SUCs on different basins or with varying meteorological or geological parameters.
\textit{HydroStartML} predicts steady-state initial DTWT configurations for SUCs based on geological properties, the potential groundwater recharge, and the slopes of the domain surface.
These predicted DTWT configurations can be corrected to actual steady-state conditions during the subsequent SUCs with a reduced computational effort compared to classical SUCs initialized with a less well informed DTWT configuration.
This allows for faster simulations in new regions or when exploring the parameter space and therefore, lowers the entry barrier for hydrological simulations.
\textit{HydroStartML} can pave the way for integrating machine learning with traditional simulation methods, with the goal of significantly increasing computational efficiency in hydrological simulations.

We will be working on the integrated simulation platform \textit{ParFlow} (PARallel FLOW) \citep{Jones2001Advances, Ashby1996Parallel, Kollet2006Integrated, Maxwell2013TerrainFollowing}, which allows constructing physically-based hydrological models to simulate the hydrological cycle. 
Therefore, we first present the governing equations for \textit{ParFlow} that are the most relevant for us and later derive our choice of the input and output variables for our machine learning emulator in Sec.~\ref{sec:input_features}.
We then discuss the general data generation procedure in Sec.~\ref{sec:data_gen} and the architecture of the CNN \textit{HydroStartML} in Sec.~\ref{sec:arch}.
Our results section is split into two parts, in which we will test the quality of the predictions of \textit{HydroStartML}, respectively, in two controlled experiments.
First, we consider test samples and compare predicted DTWT configurations to the ground truth steady-state DTWTs in Sec.~\ref{sec:results_train}.
Second, some predicted DTWTs are applied as initial conditions for subsequent SUCs in Sec.~\ref{sec:results_SUC}.
The spin-up effort until convergence is monitored and compared to an SUC that had been initialized with the classical approach of a constant DTWT.
Lastly, we derive our conclusion on the ability of \textit{HydroStartML} to predict DTWT configurations and on the induced reduction in computational spin-up effort in Sec.~\ref{sec:label}.

\section{Methods} \label{sec:methods}
In section two, we introduce \textit{HydroStartML}, a machine-learning emulator trained to predict a steady-state DTWT configuration based on four input features that describe the physical properties of the respective domain.
The training data has been generated through computationally expensive SUCs in a previous study \citep{Yang2023CONUS2}.
The predicted DTWT configuration is then applied as the initial condition for SUCs.

\subsection{ParFlow} \label{chap:ParFlow}
As an application example, we will use the integrated simulation platform \textit{ParFlow} \citep{Kollet2006Integrated, Maxwell2013TerrainFollowing}.
An extensive review of the core functionality, capability, applications, and ongoing development of \textit{ParFlow} is discussed in \citep{Kuffour2020Simulating}.
\textit{ParFlow} simulates groundwater and surface water by solving both systems simultaneously.
It uses the surface water equations as the overland flow condition, which closes the initial value problem of variably saturated groundwater flow \citep{Kollet2006Integrated}.
Its numerical grid is directly aligned with the terrain \citep{Maxwell2013TerrainFollowing}.
\textit{ParFlow} uses advanced numerical solvers and multigrid preconditioners to allow computationally efficient usage of the parallel infrastructure of \textit{ParFlow} to solve the underlying equations \citep{Kollet2010Proof}.
Several modeling platforms exist, for example, for the contiguous United States (CONUS) (\textit{ParFlow} CONUS 1.0 \citep{Maxwell2015HighResolution} and \textit{ParFlow} CONUS 2.0 \citep{Yang2023CONUS2}).
We will be working with CONUS 2.0 (\textit{CONUS2}) in the following.

\subsection{Choice of input and outputs variables for \textit{HydroStartML}} \label{sec:input_features}
We predict the DTWT configuration based on four features: the hydraulic conductivities $k_z$, the potential groundwater recharge $R=P-ET$ (precipitation minus total evaporation and transpiration while excluding runoff) \citep{Maxwell2008Interdependence}, and the topology of the domain in the form of surface slopes in the $x-$ and $y$ directions ($S_x$ and $S_y$).
The reasoning behind this choice is explained in the following.

\textit{ParFlow} solves the mixed form of Richards' equation \citep{richards1931capillary, Celia1990General} in the three spatial dimensions, which allows it to represent three-dimensional, variably saturated, conditions:
\begin{equation} \label{eq:richards}
    S_S S_W(\psi_P) \frac{\partial \psi_P}{\partial t} + \frac{\partial \phi S_W(\psi_P)}{\partial t} = \nabla \cdot \boldsymbol{q} + q_S.
\end{equation}
The specific volumetric (Darcy) flux term $\boldsymbol{q}\; [LT^{-1}]$ relies on Darcy's law \citep{Darcy1856Fontaines}:
\begin{equation} \label{eq:darcy}
    \boldsymbol{q}=-\boldsymbol{K_S}(x) k_r(\psi_P)\nabla(\psi_P + z).
\end{equation}
These expressions depend on $S_S$ as the specific storage coefficient $[L^{-1}]$; $S_W$ is the degree of saturation $[-]$; $\psi_P$ is the subsurface pressure head $[L]$; $t$ is the time $[T]$; $\phi$ is the porosity $[-]$; $q_S $ is the general sink/source term $[T^ {-1}]$;  $\boldsymbol{K_S}$ is the saturated hydraulic conductivity tensor $[LT^{-1}]$; $k_r$ is the relative permeability $[-]$; $z$ is the elevation with the z axis specified as upwards $L$ \citep{Maxwell2013TerrainFollowing}.
For an extensive derivation of the governing equations for \textit{ParFlow}, refer e.g. to \citep{Maxwell2013TerrainFollowing}.
Since we are focusing on steady-state conditions in this work, by neglecting the compressibility term and the term describing the change in water content on the left hand side, this equation simplifies to
\begin{equation} \label{eq:richards_simpl}
    \nabla \cdot (\boldsymbol{K_S}(x) k_r(\psi_P)\nabla(\psi_P + z)) = q_S.
\end{equation}
The van Genuchten relationships \citep{VanGenuchten1980Closed} describe changes in saturation $S_W$ and permeability $k_r$ with pressure.
Since this work is aimed at computing steady-state conditions, these changes in saturation $S_W$ and permeability $k_r$ are less significant for us.
In contrast, the hydraulic conductivity tensor $\boldsymbol{K_S}$ has a non-negligible impact on the Darcy flux.
Specifically, the horizontal component of the saturated hydraulic conductivities, that we will call $k$ in the following, cannot be neglected, as it is much more relevant than vertical resistance due to the grid cells being much wider than they are high.
Therefore, the horizontal hydraulic conductivities $k$ are the obvious choice for one input feature of \textit{HydroStartML}.
Also, since fluxes are factored in as a source term in the Richards' equation \eqref{eq:richards}, we consider the potential groundwater recharge $R=P-ET$ as one input feature for our model.
Lastly, we add the slopes of the surface $S_x$ and $S_y$ because we presume they might be deciding input features for our model as they allow conclusions on the interplay between neighboring grid cells.

Also, we must specify the output of \textit{HydroStartML}.
\textit{ParFlow} operates on three-dimensional pressure heads, which can be converted to a DTWT configuration. 
\textit{HydroStartML} is trained on the DTWT configurations, and this estimation is later transformed back to a pressure head configuration for applying it as an initial condition on \textit{ParFlow}.
This conversion is achieved by assuming hydrostatic equilibrium at the center of the grid cell:
\begin{equation} \label{eq:trafo_DTWT_press}
    p_i = d - \sum_{j=0}^{i-1} d_j - \frac{d_i}{2} - DTWT,
\end{equation}
where $p_i$ is the pressure head at the i'th grid cell $[L]$; $i$ is the index of the current grid cell, from bottom to top $[-]$; $d$ is the overall height $[L]$ (here: $392 m$); $d_i$ is the height of the i'th-grid cell $[L]$ (here: $200, 100, 50, 25, 10, 5, 1, 0.6, 0.3,$ $0.1 m$ for the respective layer); $DTWT$ is the DTWT, obtained from \textit{HydroStartML} $[L]$.

On \textit{CONUS2}, vertical inhomogeneity is represented with six sub-soil layers, that lie below the top four soil-layers, such that the horizontal hydraulic conductivities $k$ are represented on a three-dimensional grid.
We reduce this to a two-d grid by using the harmonic geometric average to compute the effective hydraulic conductivity in horizontal direction $k_z$ at every lateral and longitudinal position \citep{woessner2020hydrogeologic}.
This hydraulic conductivity $k_z$, together with the potential recharge $R$ and the surface slopes $S_x$ and $S_y$ are our four input features.

\subsection{Data Generation} \label{sec:data_gen}
To obtain steady-state DTWT data for training and testing \textit{HydroStartML}, we use the \textit{ParFlow} CONUS modeling platform, a continental-scale hydrological model of the contiguous United States (CONUS) \citep{Yang2023CONUS2}.
Specifically, we use existing steady-state configurations from SUCs of \textit{ParFlow} \textit{CONUS2}, which extends to the coastlines and contributing basins of the CONUS \citep{Yang2023CONUS2}.

We divide the entire CONUS2 area into patches with a spatial extent of $0.392 km \times 150 km \times 150 km$ (depth, height, width) each, which corresponds to $10 \times 150 \times 150$ grid cells.
We move across the entire \textit{CONUS2} domain in increments of $50$ cells in the lateral and longitudinal direction respectively.
As we move through the \textit{CONUS2} domain, we accept every $150 \times 150$ patch  as a data sample that consists of more than $50\%$ land.
In non-eligible regions inside those patches, e.g. those that align with the ocean or a lake, the respective values of the input features are set to zero.
Overall, this data generation procedure results in a total of $3024$ eligible data samples, that consist of the two-dimensional ground truth DTWT configuration as the output feature and four two-dimensional input features, all of size $150 \times 150$ grid cells.
We compare these data samples concerning how typical their DTWT configuration appears and exclude the eight samples for which the DTWT deviates most from the overall average.
These atypical patches will later be used to test the behavior of \textit{HydroStartML} on especially challenging regions.
The remaining $3016$ samples are randomly split into $2262$ samples used for training and $302$ samples used for validation, such that $452$ samples remain for testing purposes, which we will call testing dataset in the following.

\begin{figure}
    \centering
    \includegraphics[width=0.99\linewidth]{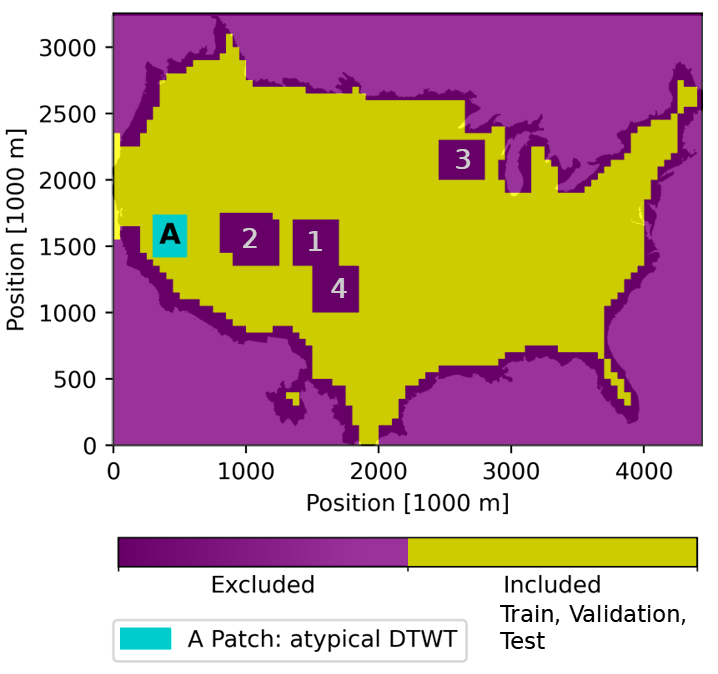}
    \caption{All patches of size $150\times150$ grid cells on the \textit{CONUS2} domain that were included and later used as training, validation and testing dataset are marked in yellow. The outline of the full \textit{CONUS2} domain is marked \citep{Yang2023CONUS2}. Excluded regions either do not belong to the CONUS, mainly consist of water or are in local proximity to the demo basins $1,2,3,4$.
    A patch with a particularly atypical DTWT (marked with an $A$) contains an atypical basin that we will use as additional demo basin.}
    \label{fig:data_gen}
\end{figure}

\begin{figure}
    \centering
    \includegraphics[width=0.99\linewidth]{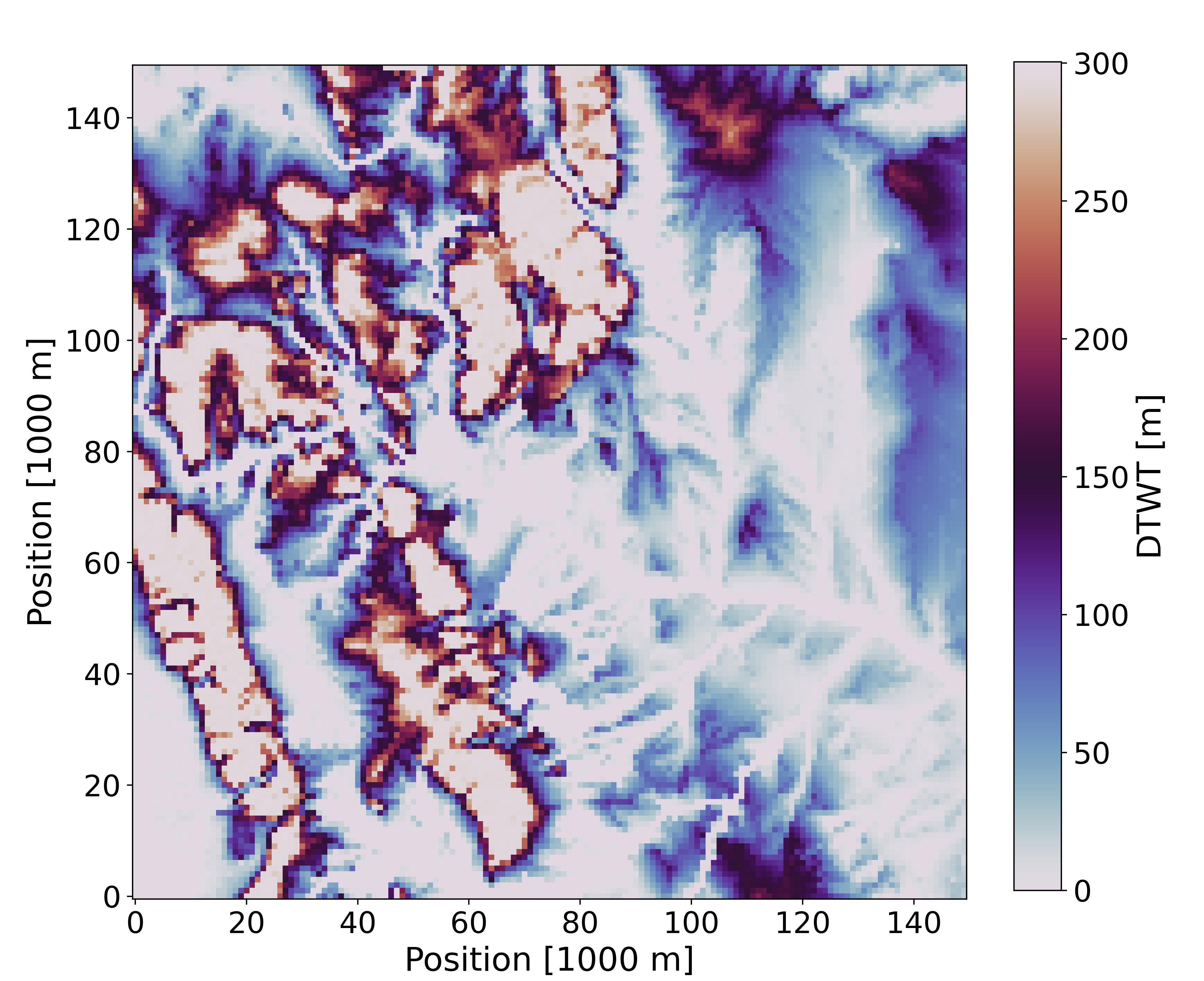}
    \caption{Ground truth steady-state DTWT configuration in a patch of $150 \times 150$ grid cells, in which demo basin $1$ is encapsulated.}
    \label{fig:demo_DTWT}
\end{figure}

In the first experimental setup (experiment \#1), we compare the predicted DTWT configurations with the ground truth steady-state DTWTs.
We conduct those experiments on the testing dataset.
With the second experiment (experiment \#2), we investigate the impact of applying some predicted DTWTs as initial conditions for subsequent SUCs to demonstrate the potential of our approach of reducing the computational effort for those.
The spin-up effort until convergence is monitored and compared to SUCs that had been initialized with different initial conditions, e.g. using the classical approach of a constant DTWT.
Therefore, we excluded those regions that contain four demo basins that we selected in advance and a sufficiently large offset from the data generation procedure.
This means that the according regions are not part of the initial dataset and therefore do not appear in the training, validation, or testing dataset.
Additionally, we found one further demo basin that is completely encapsulated in one of the eight atypical patches.
On this atypical demo basin $A$, we will test the spin-up behavior under extreme conditions.
All accepted data samples, as well as the locations of those demo basins $1$ through $4$ and $A$ and the respective surrounding regions that we excluded, are represented in Fig.~\ref{fig:data_gen}.
Exemplarily, the ground truth steady-state DTWT configuration in a patch of $150 \times 150 km$, i.e. $150 \times 150$ grid cells of demo basin $1$ is shown in Fig.~\ref{fig:demo_DTWT}.

\subsection{Architectures} \label{sec:arch}
\begin{figure}
    \centering
\includegraphics[width=0.99\linewidth]{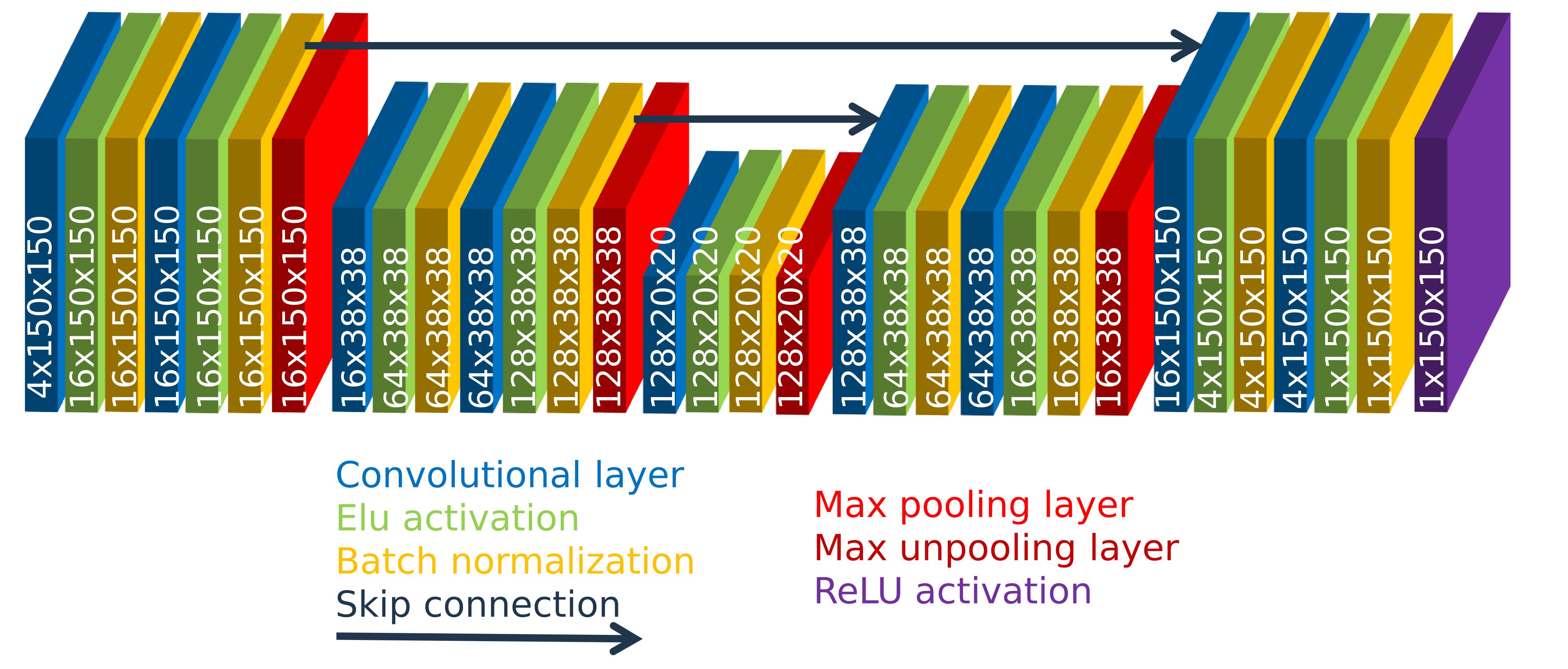}
    \caption{Architecture of the convolutional neural network \textit{HydroStartML} with a U-Net structure, including the size of the input of each layer.}
    \label{fig:arch}
\end{figure}

\textit{HydroStartML} is a convolutional neural network (CNN). 
Its architecture, shown in Fig.~\ref{fig:arch}, is loosely inspired by the well-known two-piece U-Net structure, consisting of the encoder and the decoder \citep{ronneberger2015u}.
U-Nets are popular due to their ability for capturing high-level features while still being able to precisely generate fine-grained details.
The core building blocks of both pieces are several two-dimensional convolutional layers (blue color in Fig.~\ref{fig:arch}), respectively, followed by an exponential linear unit (ELU) activation function  (green color in Fig.~\ref{fig:arch}) \citep{clevert2015fast}.
Convolutional layers allow capturing the spatial context for each grid cell and we can evaluate the ELU activation function fast while not risking dying neurons.

In the encoder, the size of the input is successively reduced (see sizes $150 \times 150$, $38 \times 38$, $20 \times 20$ in Fig.~\ref{fig:arch}) while the channel width grows (see channel widths $4$, $16$, $64$ and $128$ in Fig.~\ref{fig:arch}).
The input data is therefore transformed into a hierarchical data structure where each channel can represent a specific feature.
In \textit{HydroStartML}, the spatial extent is reduced twice.
For that, we use max-pooling layers (red color in Fig.~\ref{fig:arch}), which take the maximum of the entries that are contained with respect to a quadratical max-pooling kernel.
This kernel traditionally does not overlap and functions as a mask \citep{jarrett2009best}.
Max-pooling allows us to capture the most distinctive characteristic within this mask while at the same time introducing additional non-linearity in the network, a necessary requirement for deep neural networks.

The two pooling layers are applied respectively after two convolutional layers with the according activation function layer.
In the decoder, these max-pooling steps are symmetrically reversed with max-unpooling layers to regain an output of the same lateral and vertical extent as the input.

Additionally, we introduced batch normalization after every activation function (yellow color in Fig.~\ref{fig:arch}) because it drastically accelerated the training by fixing the mean and variances of the layer output in a normalization step, which is then scaled and shifted with learnable parameters \citep{IoffeS15}.
To make the model output oblige physical laws, we added one final ReLU layer \citep{arora2016understanding}, as we do not want to allow negative DTWTs which are therefore set to $0.0$.
Overall, \textit{HydroStartML} has $1,336,579$ learnable parameters.

To fight a vanishing gradient, we introduced two skip connections that respectively add the absolute values from the encoder to the according layer with the same size in the decoder, respectively after both max-unpooling steps.

We introduced normalization of the input features to the $[0,1]$-range, to ensure that training features with large absolute values do not outweigh other training features.
After initial tests, we rejected normalizing the output, i.e., the DTWT; we observed a decrease in performance since the calculated losses with normalized outputs were too small and hindered the optimization procedure.

\textit{HydroStartML} employs minibatching to avoid GPU overflow; for every epoch we process minibatches of a specified size that consist of random combinations of the training data set until the entire training dataset is seen once.

The underlying optimization problem for the learnable parameters is solved iteratively with a stochastic gradient descent (SGD) optimizer \citep{lecun1998gradient}. 
The model weights are initialized with the commonly used Xavier uniform distribution \citep{Glorot2010WeightInit}, and the biases are set to $0.01$.
The SGD optimizer is applied with a learning rate of $10^{-4}$, a momentum of $0.9$, and a Nesterov momentum.
Due to the learning rate scheduler, the initial learning rate of $10^{-4}$ decays exponentially with a multiplicative factor of $0.85$. 
Early stopping is activated if the observed validation loss exceeds the overall ever observed minimum by more than a previously determined threshold for $20$ times in a row.

We determined the optimizer hyperparameters, namely the initial learning rate, the multiplicative factor in the learning rate scheduler, and the momentum of the optimizer, during an extensive hyperparameter optimization.
The performance of the model depending on these hyperparameters is shown in Appendix \ref{app:C2_compare_archits}.

\section{Results and Discussion: Training and Testing} \label{sec:results_train}
\begin{figure}
    \centering
    \includegraphics[width=0.99\linewidth]{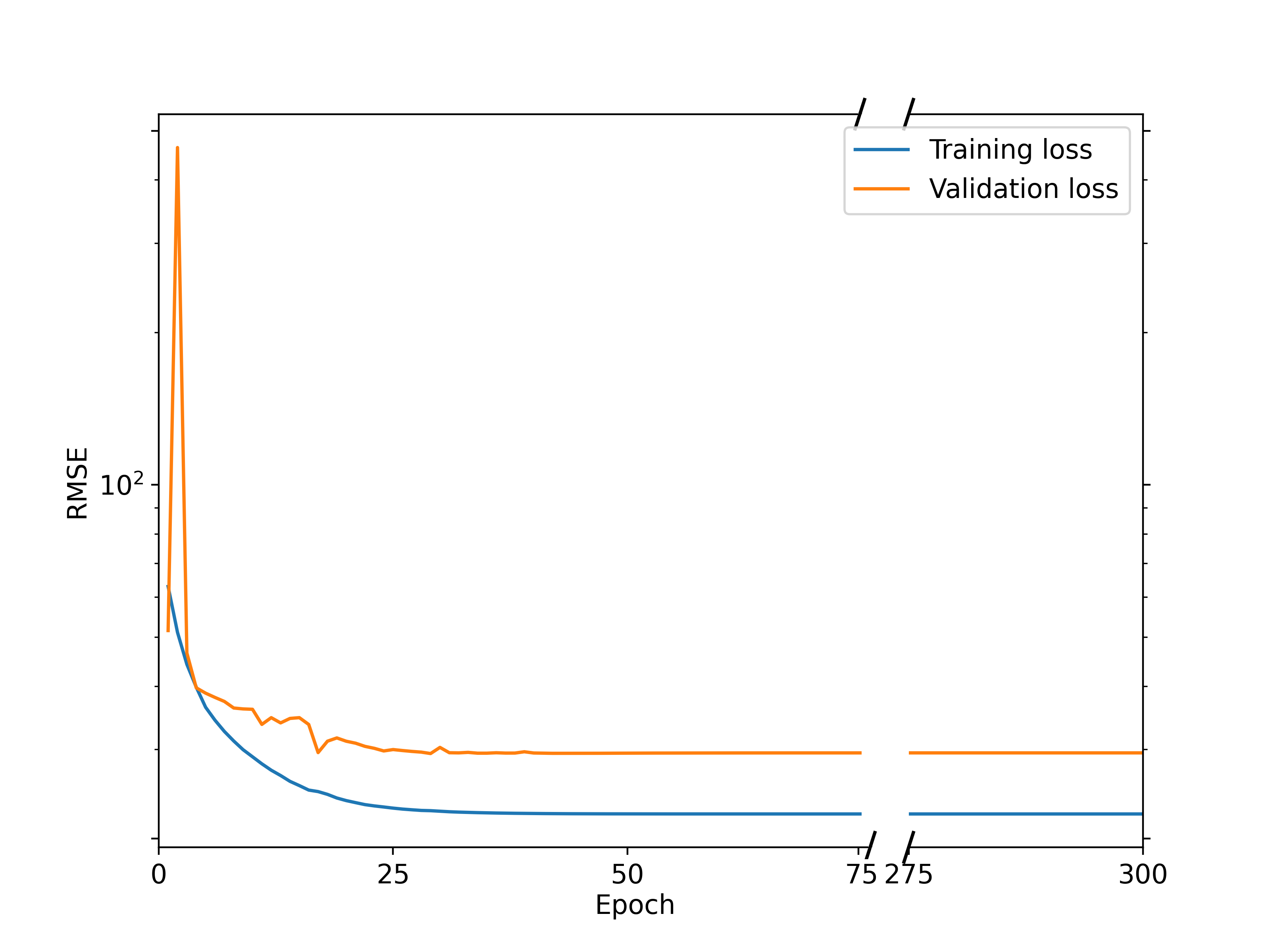}
    \caption{Development of the RMSE on the training and validation dataset during training of \textit{HydroStartML}.}
    \label{fig:C2_MSEloss}
\end{figure}

We now carry out the training of \textit{HydroStartML} by using the training dataset, before testing it on the testing dataset.
After $125$ epochs, there is no identifiable change in either the training or validation error (see the development of training and validation losses in Fig.~\ref{fig:C2_MSEloss}).

With our first experiment as introduced above (experiment \#1), we evaluate the ability of the trained \textit{HydroStartML} to predict DTWT configurations that are close to the ground truth steady-state DTWTs on the testing dataset.
The final average RMSE on a test patch of $150 \times 150 km$ is $28.67m$, or when considering the eight atypical samples as additional test samples, $29.90m$.

\subsection{Modeled DTWTs}
\begin{figure}[!ht]
    \begin{subfigure}[b]{\columnwidth}
        \begin{subfigure}[b]{0.49\columnwidth}
            \centering
            \includegraphics[height = 3.2 cm, keepaspectratio]{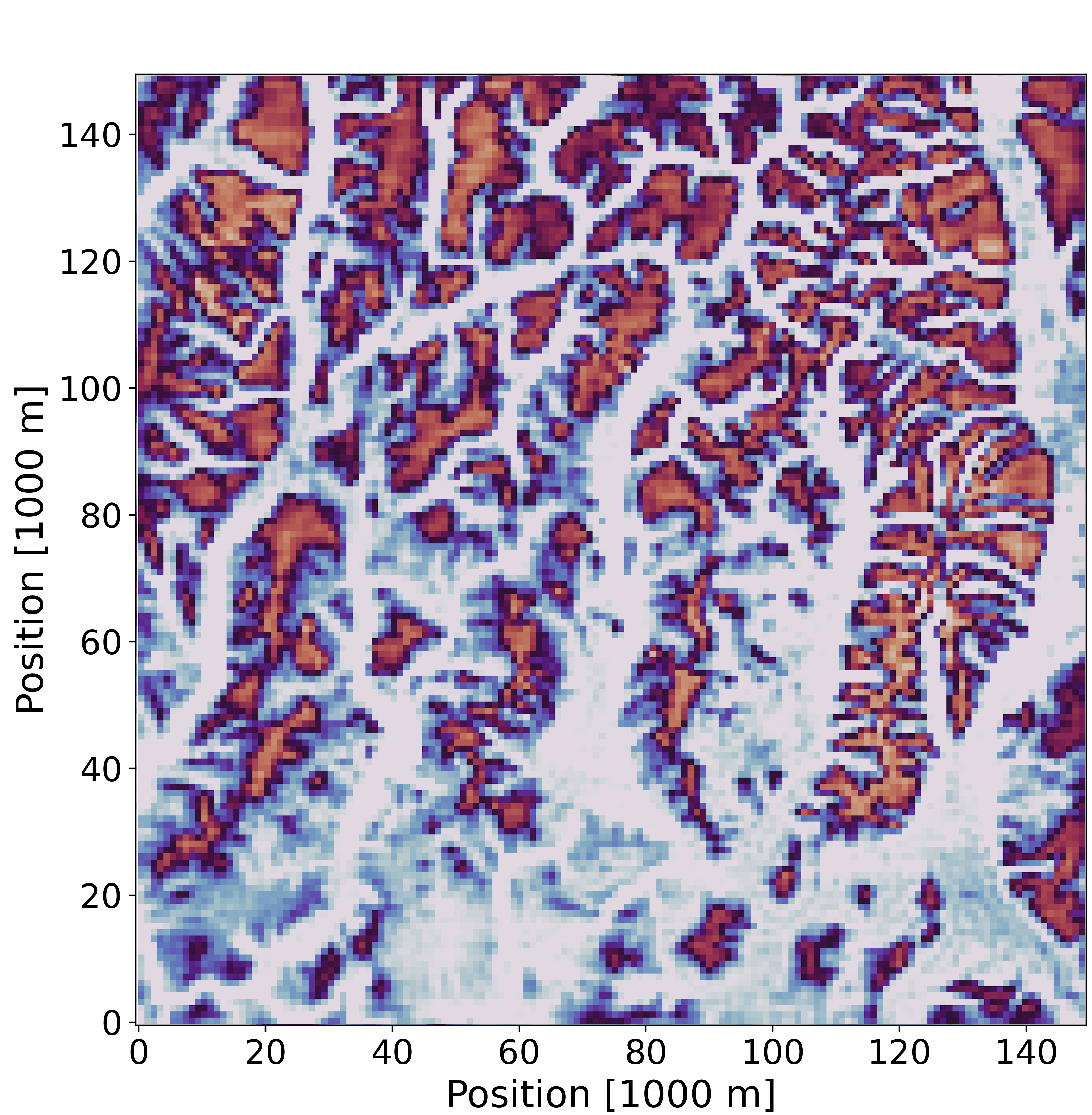}
            \caption*{i) Modeled DTWT configuration}
        \end{subfigure}
        \hfill
        \begin{subfigure}[b]{0.49\columnwidth}
            \centering
            \includegraphics[height= 3.2 cm, keepaspectratio]{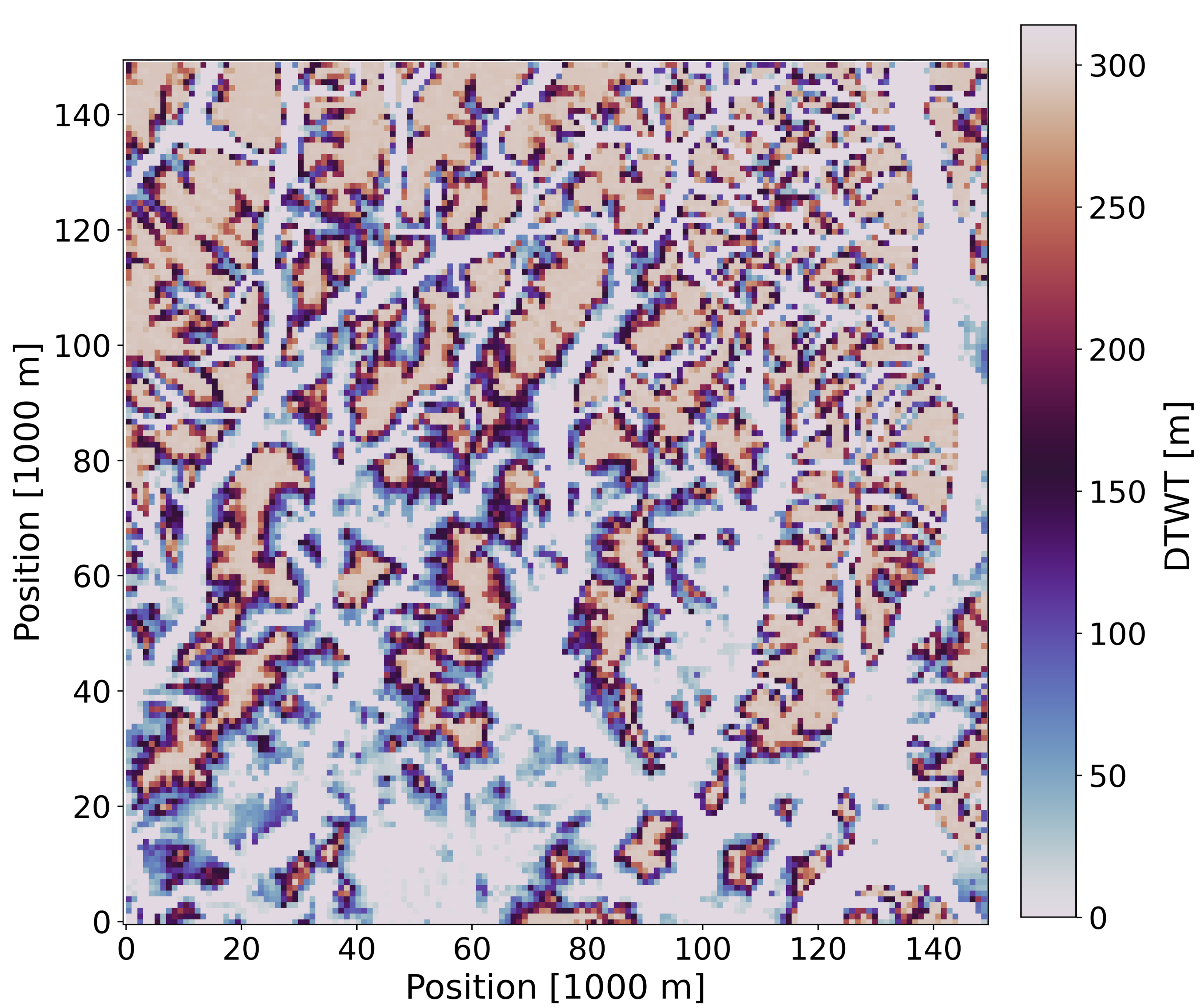}
            \caption*{ii) Ground truth}
        \end{subfigure}     
    \subcaption*{(a) i) Modeled DTWT configuration and ii) ground truth.}
    \end{subfigure}
    \caption[Modeled DTWT configurations and ground truths on test sample]{Modeled DTWT configuration and ground truth on test sample.}
    \label{fig:Mod_DTWTs_test}
\end{figure}

To visualize and discuss the DTWT configurations behind these metric values, we show the DTWT configurations that we predicted with \textit{HydroStartML} in comparison to the true DTWT configuration on three exemplary patches.
First, we show one patch from the testing dataset that had been randomly selected from the yellow region as shown in Fig.~\ref{fig:data_gen} (experiment \#1).
This particular test patch has a RMSE of $61.33 m$, which is worse than $95.3 \%$ of patch-wise RMSE values on test samples.
On this test sample, we see that the main characteristics of the ground truth DTWT map are predicted correctly by \textit{HydroStartML} (Fig.~\ref{fig:Mod_DTWTs_test}).
Generally, we observe that shallow DTWTs are represented particularly well, while the DTWTs in areas with larger depths seem to be underestimated by \textit{HydroStartML}.
This might be due to a slightly too steep decline in the learning rate.

\begin{figure}[ht!]
    \begin{subfigure}[b]{\columnwidth}
        \begin{subfigure}[b]{0.49\columnwidth}
            \centering
            \includegraphics[height= 3.2 cm, keepaspectratio]{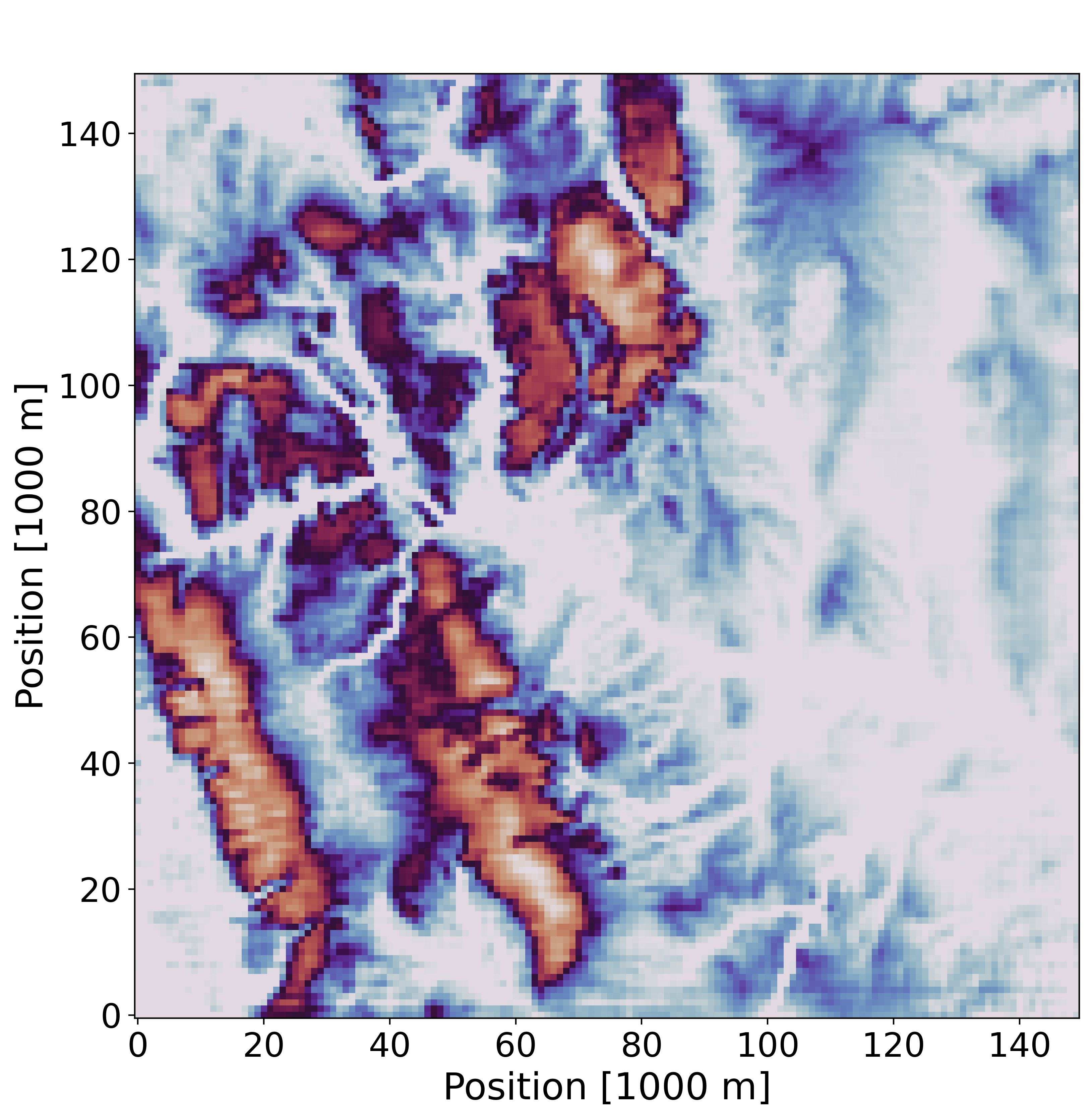}
            \caption*{i) Modeled DTWT configuration}
        \end{subfigure}
        \hfill
        \begin{subfigure}[b]{0.49\columnwidth}
            \centering
            \includegraphics[height= 3.2 cm, keepaspectratio]{figures/0209_0_11020002_estimation_3_gt.png}
            \caption*{ii) Ground truth}
        \end{subfigure}     
    \subcaption* {(b) i) Modeled DTWT configuration and ii) ground truth.}
    \end{subfigure}
    \caption[Modeled DTWT configurations and ground truths on demo basin]{Modeled DTWT configuration and ground truth on demo basin $1$.}
    \label{fig:Mod_DTWTs_demo}
\end{figure}

\begin{figure}[ht!]
    \begin{subfigure}[b]{\columnwidth}
        \begin{subfigure}[b]{0.49\columnwidth}
            \centering
            \includegraphics[height= 3.2 cm, keepaspectratio]{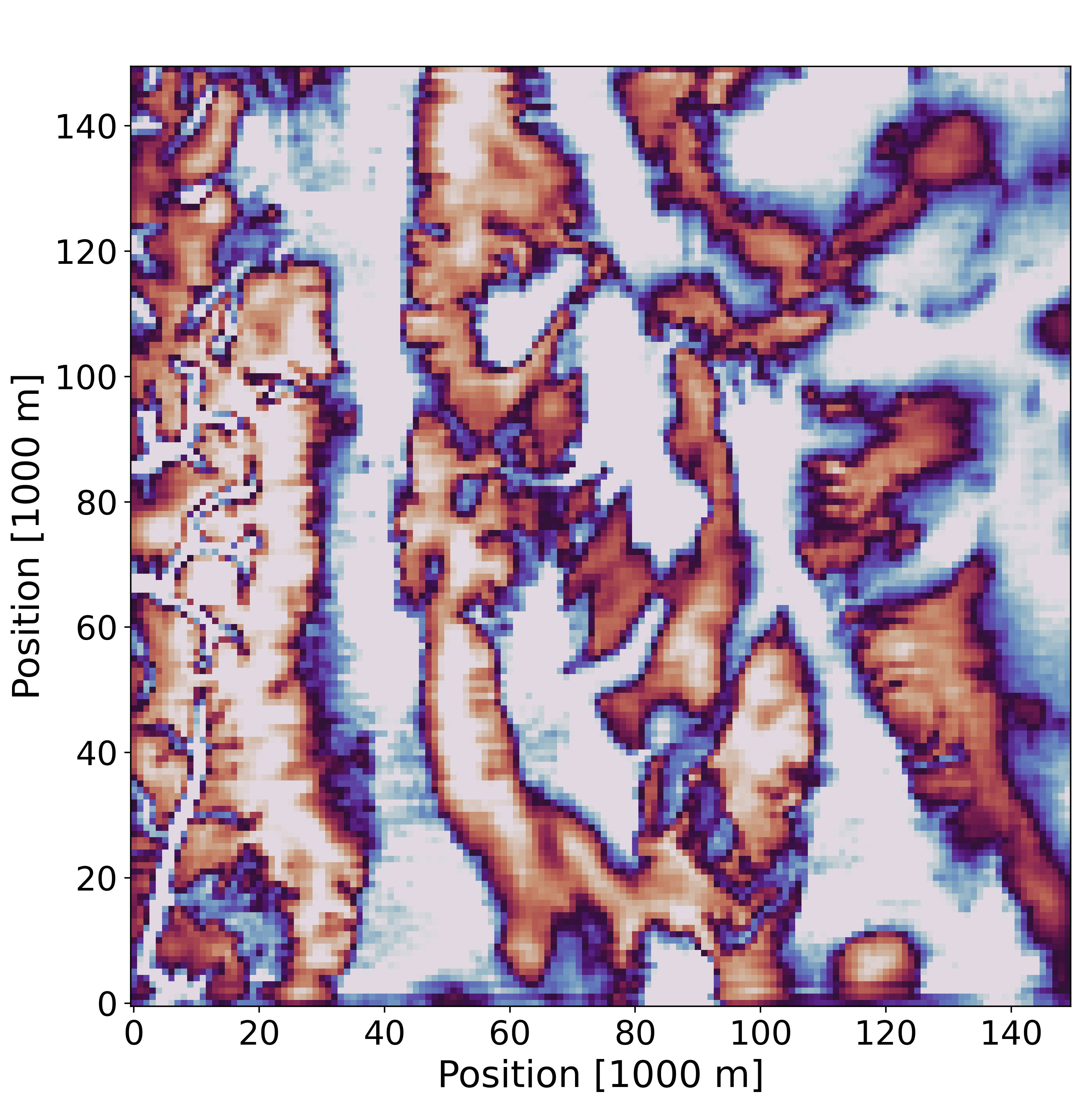}
            \caption*{i) Modeled DTWT configuration}
        \end{subfigure}
        \hfill
        \begin{subfigure}[b]{0.49\columnwidth}
            \centering
            \includegraphics[height= 3.2 cm, keepaspectratio]{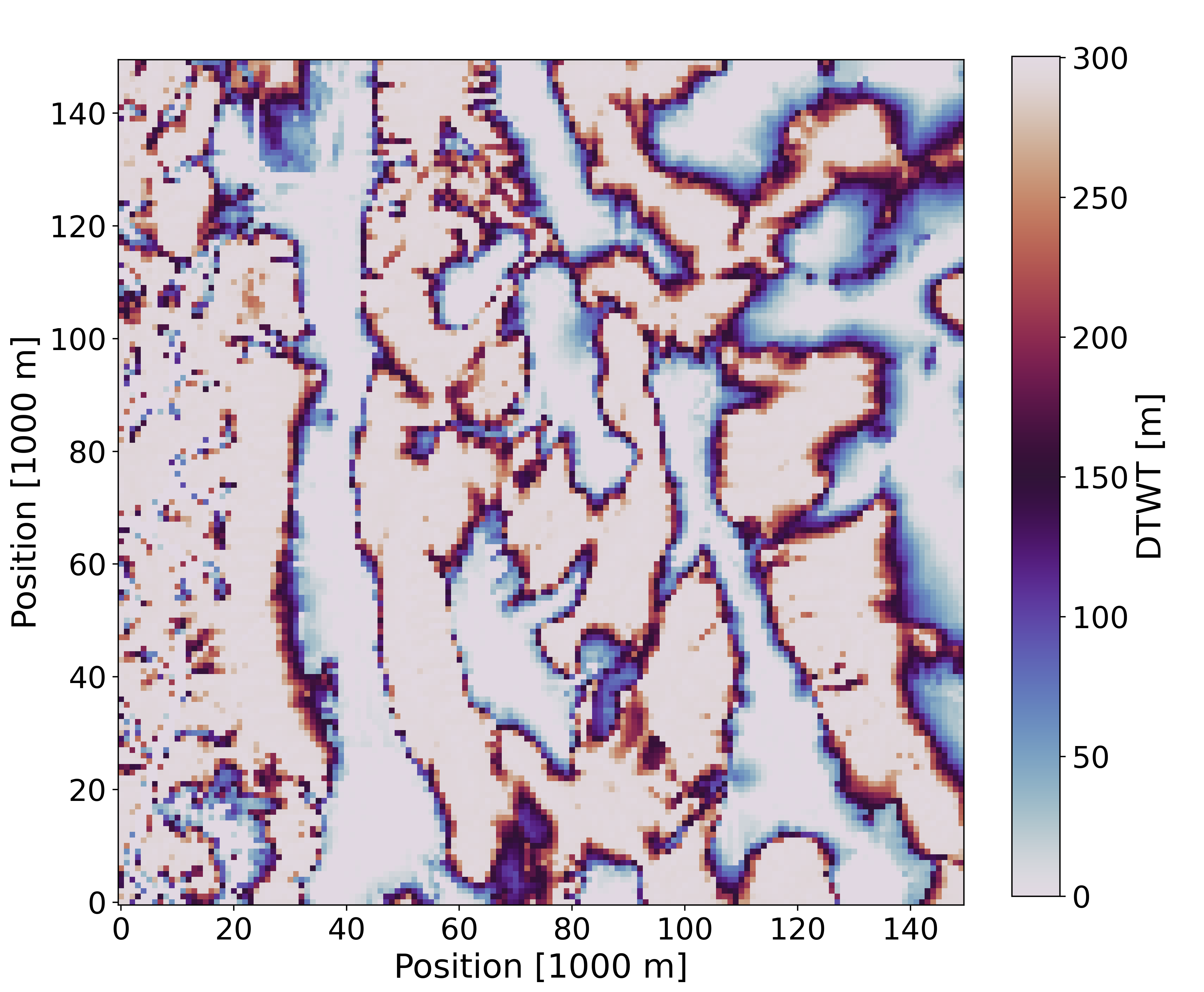}
            \caption*{ii) Ground truth}
        \end{subfigure}     
    \subcaption*{(c) i) Modeled DTWT configuration and ii) ground truth.}
    \end{subfigure}
    \caption[Modeled DTWT configurations and ground truths on atypical basin]{Modeled DTWT configuration and ground truth on atypical basin $A$.}
    \label{fig:Mod_DTWTs_atypical}
\end{figure}

Second, we look at two patches on which we will investigate the spin-up behavior in experiment \# 2 in Sec.~\ref{sec:results_SUC}.
We show the demo basin number $1$ (marked with a $1$ in Fig.~\ref{fig:data_gen}) in Fig.~\ref{fig:Mod_DTWTs_demo}, which was predicted with  RMSE of $35.40 m$, which is better than $26.5 \%$ of all test samples.
Also, we investigate one of the previously excluded atypical patches that contains the later investigated atypical basin (marked with $A$ in Fig.~\ref{fig:data_gen}, shown in Fig.~\ref{fig:Mod_DTWTs_atypical}), whose DTWT was predicted with a RMSE of $69.15 m$, which is worse than $98.2 \%$ of all test samples.
These two patches confirm our previous observation that deviations mostly occur in the regions with deep water tables, whereas areas with shallow water tables seem to be represented better.


\subsection{Behavior on Test Data and on Atypical Patches}
\begin{figure}[!ht]
    \begin{overpic}[width=\linewidth]{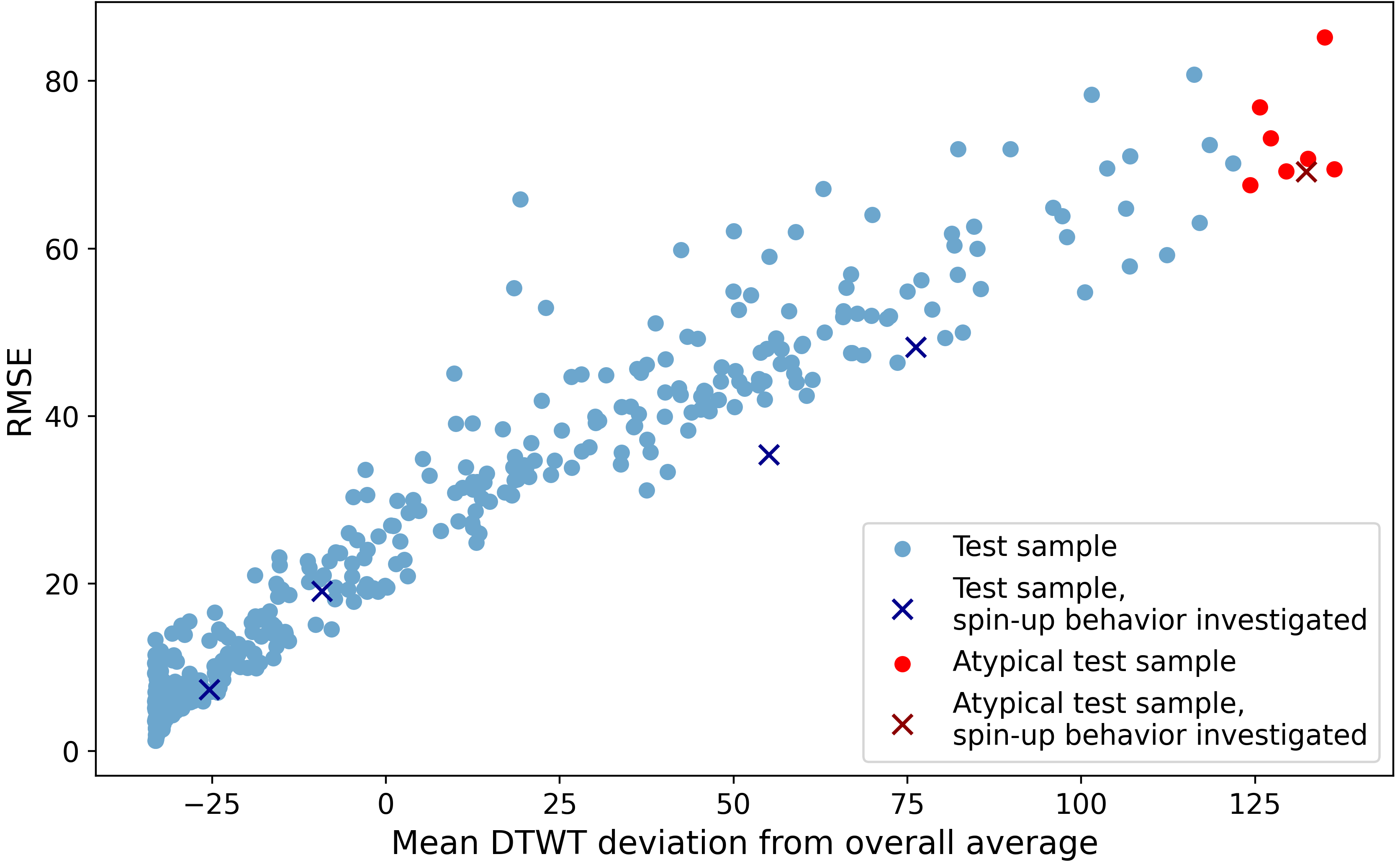}
    \put(56,28){\color{blue}1}
    \put(65,32.5){\color{blue}2}
    \put(15,8.5){\color{blue}3}
    \put(23,13){\color{blue}4}     
    \put(92,44.5){\color{red}A}
    \end{overpic}
    \caption[RMSE on test samples on \textit{CONUS2} over how typical each test sample is]{RMSE of test samples over how typical each test sample is, computed as mean deviation of the ground truth DTWT configuration of this test sample from the overall average DTWT (negative values correspond to a more shallow DTWT than on average). The RMSEs on the demo basins $1$ through $4$, as well as the atypical patch on which the spin-up behavior is investigated in experiment \#2, are marked.}
    \label{fig:C2_devi_vs_loss}
\end{figure}
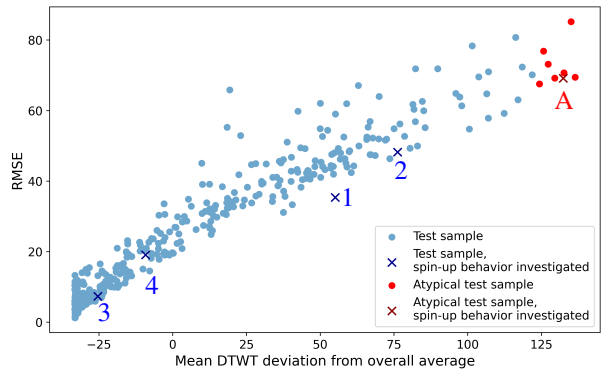

To further analyze the composition of the overall DTWT prediction quality on the testing dataset, we investigate the relationship of the patch-wise RMSE on all test samples compared to how ``atypical'' each DTWT configuration of the respective test sample is. 
From that, we hope to learn in which settings \textit{HydroStartML} can be expected to perform particularly well.
The results are shown in Fig.~\ref{fig:C2_devi_vs_loss}.

The ``atypicality'' on the $x-$ axis is defined as the difference between patch-wise DTWT of each testing sample from the global, \textit{CONUS2-} wide averaged DTWT.
This overall averaged DTWT represents the average across all training, testing and validation patches from the entire \textit{CONUS2} domain (represented in yellow in Fig.~\ref{fig:data_gen})
It is almost uniformly distributed and behaves when used as the initial condition for SUCs comparable to the constant initial condition, but has an average value of $32.26m$.

As can be seen in Fig.~\ref{fig:C2_devi_vs_loss}, DTWT configurations that are, on average, more shallow than this overall average are predicted with a low error by \textit{HydroStartML}.
Deeper DTWT configurations are, in general, predicted with a larger error.
Apparently, the less shallow the DTWT configuration is, the more difficult it is to be predicted by \textit{HydroStartML}.

\subsection{Feature Importance}
As just seen, \textit{HydroStartML} seems to generally be able to predict DTWT configurations that capture the main characteristics of the ground truth DTWTs.
We now investigate the relative importance of the four input features, namely the hydraulic conductivities $k_z$, the potential groundwater recharge $R$, and the terrain in the form of the slopes of the surface $S_x$ and $S_y$.

\begin{figure}
    \centering
    \includegraphics[width=0.99\linewidth]{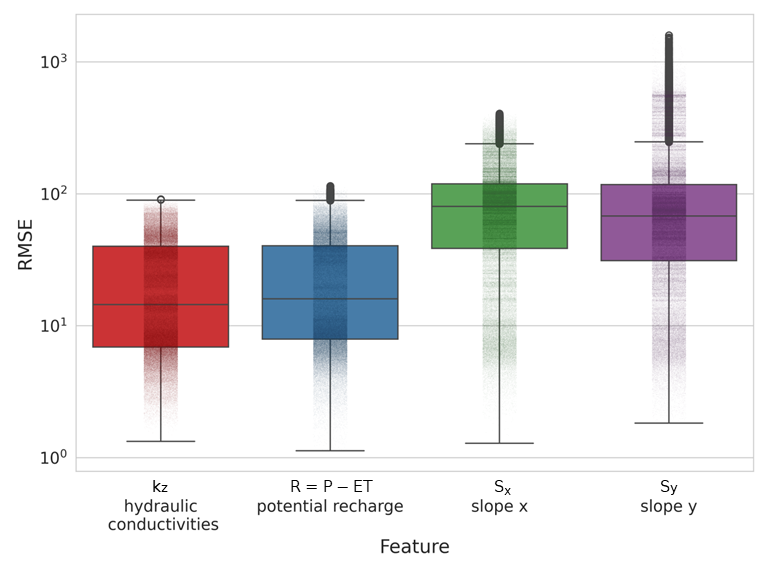}
    \caption{Feature importance of the four input features of \textit{HydroStartML}: the hydraulic conductivities $k_z$, the potential groundwater recharge $R=P-ET$, and the terrain in the form of the slopes of the surface $S_x$ and $S_y$.
    The feature importance is evaluated as the error when randomizing the respective input feature.}
    \label{fig:feat_imp}
\end{figure}

Generally, a feature that is important for the behavior of a machine learning emulator has a sensitive impact on the resulting error on test samples.
We permute one input feature (e.g. $k_z$) by swapping this input feature $k_z$ out with the $k_z$ from a different test sample, query \textit{HydroStartML} to predict a DTWT configuration based upon these new input features, and evaluate the resulting RMSEs.
We do so for all $452$ test samples, respectively for all four input features.
The RMSE induced by randomizing that feature signifies the importance of that feature.
Fig.~\ref{fig:feat_imp} shows the results for all four input features as the minimum and maximum observed RMSE across the $452 \cdot 451$ permutations for all $4$ input features, the quantiles of the error, and outliers.

The large errors when permuting the slopes of the test samples show clearly that the terrain in the form of the surface slopes is by far the deciding input feature.
The importance of the slopes $S_x$ and $S_y$ appear similar, even though more extreme outliers occur when permuting the slopes in $y-$direction.
The less important hydraulic conductivity and potential recharge seem to be similar in relative feature importance.

\section{Results and Discussion: Impact on Spin-up Effort} \label{sec:results_SUC}
In this section, we will answer our main research question: to which extent initializing an SUC with a DTWT configuration found by \textit{HydroStartML} actually reduces the overall computational effort in SUCs until a steady-state configuration is found (experiment \#2).  
To visualize and define the convergence in SUCs, we evaluate the change in total water storage inside the basin per spin-up timestep, normalized with the potential recharge.
One timestep consists of simulating $5000$ hours under constant atmospheric settings.
A well-converged system would show little to no change in total water storage per computing step.

For comparison, we evaluate the convergence behavior with several initial conditions apart from the initial condition proposed by \textit{HydroStartML} to generate a benchmark.
We do so on the four demo basins (marked with numbers $1$ - $4$ in Fig.~\ref{fig:data_gen}) that we previously omitted from training, validation, and testing in Sec.~\ref{sec:data_gen}.
Before using them as initial conditions in SUCs, DTWTs must be converted into pressure heads on the full $3-D$ grid of \textit{ParFlow}. To this end, the DTWT configurations are cropped to the bounds of the respective demo basin and converted to a pressure head by assuming hydrostatic equilibrium at the center of each grid cell.
For comparison, we do so with four different initial conditions: 1) The \emph{benchmark} DTWT configuration from \citep{Yang2023CONUS2} that was used as ground truth to train \textit{HydroStartML}, 2) a constant DTWT of $20 m$ across the entire basin, 3) the \textit{CONUS2}-wide averaged DTWT of $32.26 m$, and 4) the DTWT configuration predicted by \textit{HydroStartML}.
We consider the benchmark DTWT configuration as optimal because this configuration corresponds to what \textit{HydroStartML} is trained to predict.

\begin{figure}[ht!]
    \centering
    \includegraphics[width=0.95\linewidth]{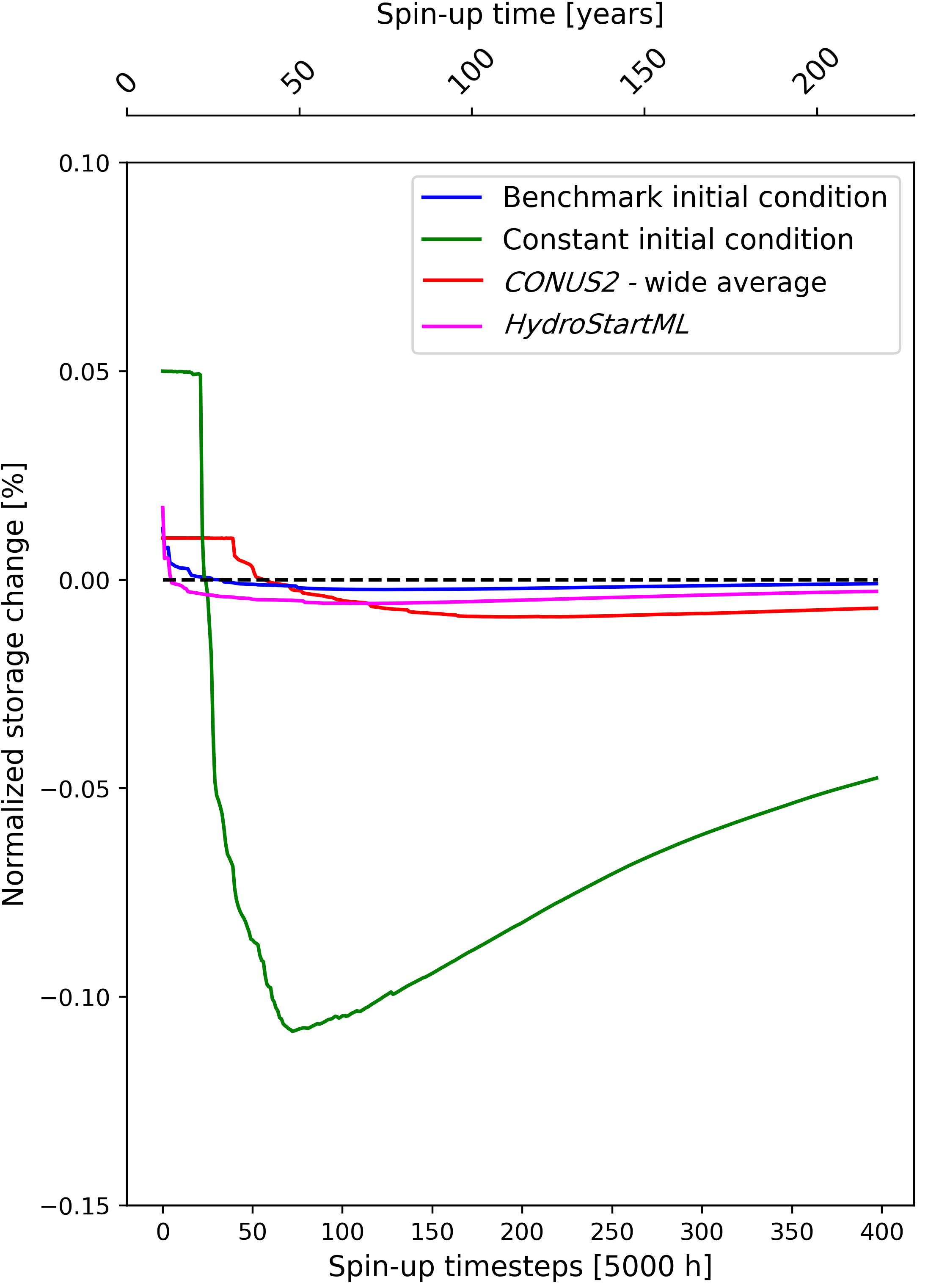}
    \caption{Convergence behavior of the normalized storage change during SUCs on demo basin $1$, initialized with different initial DTWT configurations.}
    \label{fig:spin-up-B1}
\end{figure}

To enter the discussion, we exemplarily show the spin-up behavior on demo basin $1$ in Fig.~\ref{fig:spin-up-B1} as an explicit example.
First, the benchmark initial condition (blue line) from \citep{Yang2023CONUS2} converges almost immediately against a storage change of zero, as expected.
Second, the \textit{CONUS2-}wide average DTWT (red line) plateaus for the first $50$ timesteps.
During those initial spin-up timesteps, we observe a redistribution of the internally stored water while the added potential recharge continuously is absorbed.
This is reflected in a constant storage change, which does not further converge at the time scales investigated here.
SUCs initialized with a constant DTWT of $20m$ (green line) show a similar plateau at a higher normalized storage change level.
After some initial time steps, the additionally absorbed water, as well as the excess water due to the more shallow initialization of the DTWT, drains.
This leads to a large, negative storage change with a typical exponential-asymptotic (but very slow) approach to steady state that clearly shows no satisfactory state in the investigated time frame.
Finally, the SUCs using the initial condition as predicted by \textit{HydroStartML} behave similarly to the SUCs initialized with the benchmark DTWT.
This behavior is to be expected and desirable, as it shows that \textit{HydroStartML} can predict helpful DTWT configurations that are close to the ground truth.
This makes \textit{HydroStartML} especially valuable since the ``benchmark'' configuration is usually not available in application for initializing an SUC, as it already results from SUCs.

\begin{figure}
    \centering
    \includegraphics[width=0.99\linewidth]{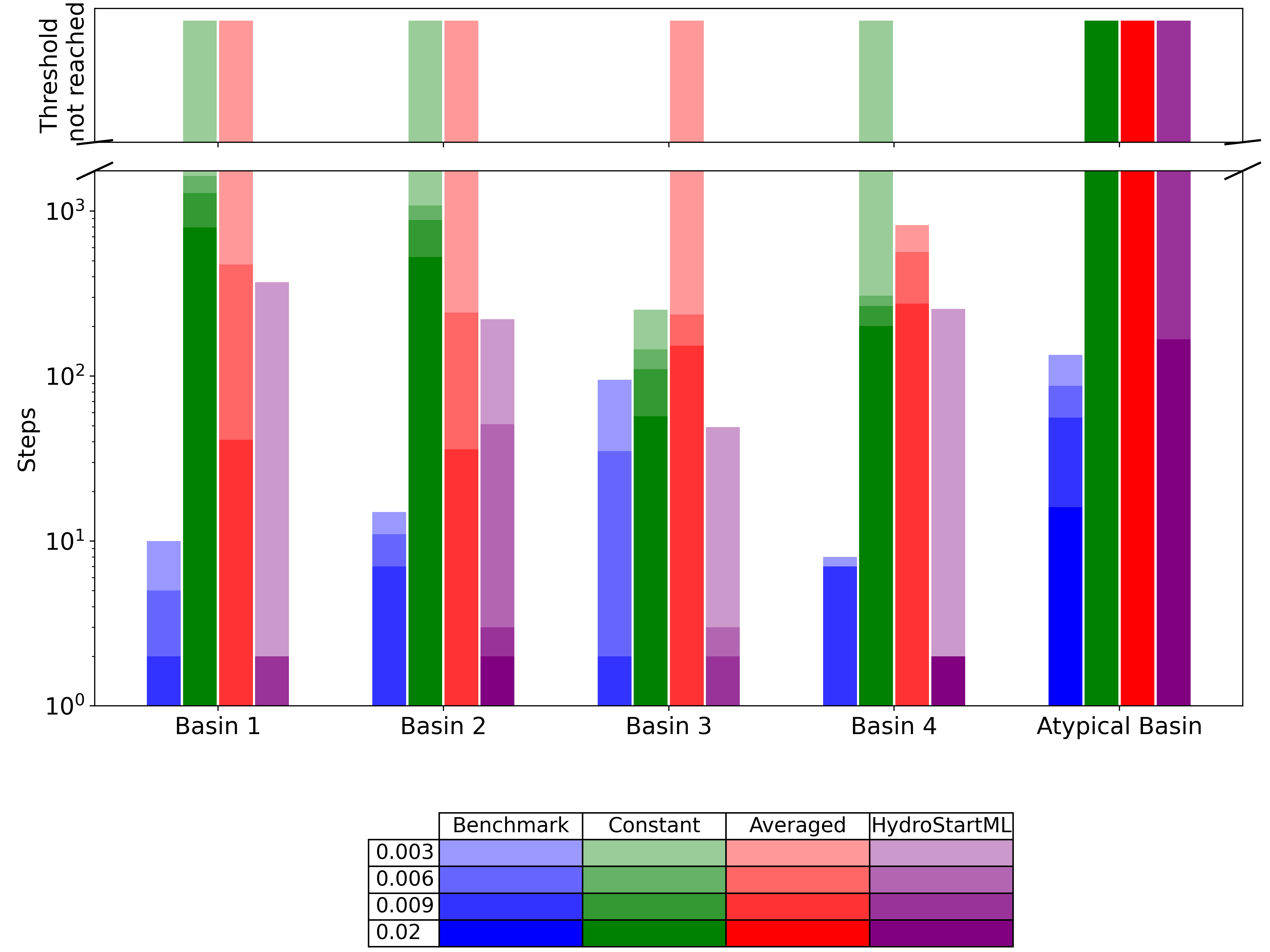}
    \caption{Necessary spin-up timesteps of respectively $5000h$ to reach different storage change thresholds when initializing an SUC with different initial DTWT configuration. We compare the initial condition predicted by \textit{HydroStartML} with the benchmark initial condition, a constant DTWT configuration, and the \textit{CONUS2}- wide average DTWT.}
    \label{fig:barplot_savings}
\end{figure}

To provide quantitative results across all four demo basins, we compare the performance of the SUCs based on four different storage change thresholds: $0.02 \%, 0.009 \%, 0.006 \%$, and $0.003 \%$.
Fig.~\ref{fig:barplot_savings} shows the number of necessary spin-up timesteps to reach these thresholds for each initialization method.
On all demo basins, the initial DTWT configuration predicted by \textit{HydroStartML} performs better than both the classical approach of a constant DTWT of $20m$, as well as the \textit{CONUS2}-wide average DTWT as the more advanced initial condition.
It often behaves even comparable to the benchmark solution and is able to reach a convergence of the strictest threshold in the demo basins $1-4$.

On the atypical basin, SUCs initialized with \textit{HydroStartML} are the only ones apart from the benchmark solution that are able to reach at least the first convergence threshold.
The spin-up behavior when applying the initial condition predicted by \textit{HydroStartML} is by far the closest to the behavior when using the benchmark initial condition and, therefore, also the closest to convergence.
Therefore, this test of \textit{HydroStartML} in this extreme setting indicates that the presented approach is suitable to reduce the computational effort and enable faster SUCs even in very challenging terrain.

\section{Conclusion} \label{sec:label}
Equilibrium depth-to-water table (DTWT) configurations are necessary as initial conditions for hydrological simulations.
Often times, huge computational efforts are necessary to conduct SUCs until equilibrium.
Therefore, the spin-up effort significantly impacts the overall computational resources necessary for any hydrological simulation.
This is especially true when no initial steady-state DTWT configuration is available in advance, such that the spin-up computations are initialized with a constant DTWT.
The observed spin-up effort was particularly substantial in basins in regions with deep DTWTs like the western US when initializing with constant DTWTs of $20m$.
The computational effort can be reduced by an informed choice of the initial condition, as traditional initialization techniques can result in an extra computational overhead.

The steady-state DTWT configurations found with spin-up computations on \textit{ParFlow} depend on various physical parameters of the considered basin, among others, the topography, represented by the surface slopes $S_x$ and $S_y$, the geological parameters as e.g. the hydraulic conductivities $k_z$, and the potential groundwater recharge $R=P-ET$.
These physical parameters are, when serving as input features, sufficient to train our CNN called \textit{HydroStartML} to function as a machine learning emulator that is able to predict a steady-state DTWT configuration.

These predicted DTWT configurations are able to capture the main characteristics of the true DTWT configurations, such that \textit{HydroStartML} is able to provide meaningful steady-state DTWT configurations on unseen terrain.
To this end, the topography of the terrain, i.e., the surface slopes $S_x$ and $S_y$, are the most important feature for successfully generating a DTWT configuration that closely matches the ground truth.

\textit{HydroStartML} can provide an initial condition for spin-up computations on unseen terrain that significantly reduces the computational effort to reach steady-state during SUCs.
This holds true even in highly complex and atypical DTWT configurations. 
\textit{HydroStartML} outperformed all other investigated initialization methods for SUCs and led to a significant reduction in computational effort compared to the classical approach of initializing the DTWTs with constant $20m$.

\textit{HydroStartML} allows this reduction in computational effort while requiring no additional knowledge, e.g. on the average DTWT in the investigated region.
As the input features for \textit{HydroStartML} need to be gathered for the subsequent hydrological simulation with \textit{ParFlow},  \textit{HydroStartML} can be applied to lower the entry barrier for any hydrological simulation on unknown or complex terrain on basins that can be confined within $150 \times 150$ grid cells.

Nevertheless, we so far only tested \textit{HydroStartML} on a limited number of demo basins which all lie inside the \textit{CONUS2} domain. 
Therefore, further investigations on its generalization abilities on unseen terrain, e.g. on a different continent, and on basins that are e.g. closer to the equator or lie on the southern hemisphere, may be informative and teach us more about the limitations of our approach.
Also, the predictions by \textit{HydroStartML} especially are prone to error in regions with large DTWTs.
Since especially deep water tables negatively impact the spin-up effort, further investigations on how to reduce these errors e.g. by adding additional penalties during the training procedure could result in even larger reductions in computational spin-up effort.
Lastly, \textit{HydroStartML} so far is limited to regions of $150 \times 150$ grid cells.
Further research on dividing larger basins into patches of $150 \times 150$ grid cells and using several predictions of \textit{HydroStartML}, or applying the concept to larger patches from scratch, needs to be conducted in the future, as a more flexible emulator would allow the reduction of spin-up effort on basins that cannot be contained inside a region of these dimensions.

The integration of machine learning with conventional simulation methods gives \textit{HydroStartML} the potential to significantly enhance computational efficiency in hydrological simulations on new regions or when exploring the parameter space. 
This advancement supports the development of hybrid approaches that allow fast and meaningful contributions to water resource management by reducing the effort required in setting up the hydrological model.

\newpage
\appendix
\section{Hyperparameter optimization of \textit{HydroStartML}}
\label{app1}

\begin{figure}[!ht]
    \centering
    \includegraphics[width=0.95\linewidth]{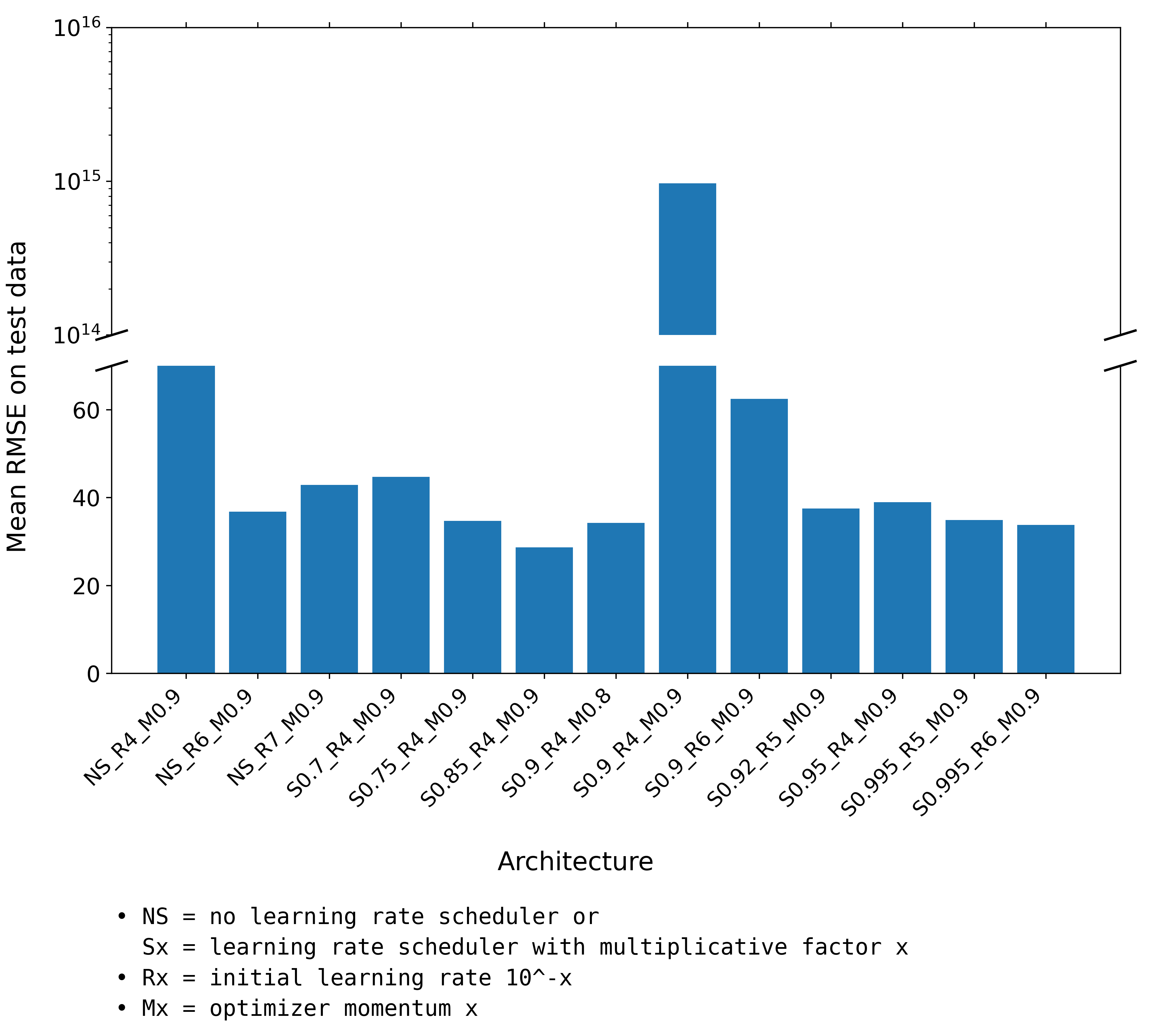}
    \caption[Performance comparison of optimizer hyperparameters on \textit{HydroStartML}]{Performance comparison of several optimizer hyperparameter settings on \textit{HydroStartML}.}
    \label{app:C2_compare_archits}
\end{figure}

\newpage
\bibliographystyle{elsarticle-num} 
\bibliography{bibliography}






\end{document}